\documentclass[fleqn]{mnras}  
\usepackage{newtxtext,newtxmath}
\usepackage[T1]{fontenc}
\usepackage{ae,aecompl}
\usepackage{graphics,graphicx}
\usepackage{amsmath}	

\newcommand{\ergs}{erg s$^{-1}$ }

\title[Implications on the scatter of the $M-T$ and $L-T$ relations]
{Low X-ray surface brightness clusters:
Implications on the scatter of the $M-T$ and $L-T$ relations
}
\author[S. Andreon et al.]{S. Andreon$^{1}$\thanks{E-mail:stefano.andreon@inaf.it}, G. Trinchieri$^{1}$, A. Moretti$^{1}$
\\
$^1$ INAF--Osservatorio Astronomico di Brera, via Brera 28, 20121, Milano, Italy\\
}
\date{Accepted ... Received ..; in original form ..}
\pubyear{2020}

\begin{document}
\label{firstpage}
\pagerange{\pageref{firstpage}--\pageref{lastpage}}
\maketitle

\begin{abstract} 
We aim at 
studying scaling relations of a small but well defined sample of galaxy clusters that includes the recently discovered 
class of objects that are X-ray faint for their mass. These clusters
have an average low X-ray surface brightness, 
a low gas fraction  and are 
under-represented (by a factor of 10) in X-ray surveys or entirely absent in SZ surveys. 
With the inclusion of these objects, we find that 
the temperature-mass relation has an unprecedented large scatter, $0.20\pm0.03$ dex at fixed mass, as wide as
allowed by the temperature range, and the location of a
cluster in this plane depends on its surface brightness.
Clusters obey a relatively tight luminosity-temperature
relation
independently of the their brightness. 
We interpret the wide difference in scatter around the two relations 
as due to the fact that 
X-ray luminosity and temperature are dominated by photons
coming from small radii (in particular for $T$ we used a 300 kpc aperture radius) and
strongly affected by gas thermodynamics (e.g. shocks, cool-cores),
whereas mass is dominated by dark matter at large radii.
We measure a slope of $2.0\pm0.2$ for the 
$L_{500}-T$ relation.  Given
the characteristics of our sample, this value is free from the collinearity (degeneracy) between evolution and slope and from hypothesis on the undetected population, that both affect the analysis of X-ray selected samples, and can therefore be profitably used both as reference and to break the above degeneracy of X-ray selected-samples.
\end{abstract}


\maketitle

\section{Introduction}

Under thermodynamic equilibrium condition, the temperature of an isolated isothermal sphere of ideal gas is
determined by the depth of the sphere potential well, $T\propto M^{2/3}$, also known as self-similar scaling, with no
scatter and no covariance with any other quantity,  (Kaiser 1986).
Deviations from this ideal behaviour, in particular intercept and slope of the mean
relation, the scatter around the mean relation, and its covariance with 
other parameters, can be used to gather information
about the processes that shape galaxy clusters (e.g. Yang et al. 2009). Pre-heating and cooling 
might alter the  slope of the scaling relation (Stanek et al. 2010);
cooling, star formation and feedback might alter both slope and intercept (Fabjan et al. 2010), while cooling and star formation might
introduce a scatter (Nagai et al. 2007a).  The variety of the mass accretion histories may also introduce a scatter
and an offset (Yang et al. 2009, Ragagnin \& Andreon 2022), 
and the offset from the mean relation is covariant with 
other cluster properties, such as concentration or formation redshift (Yang et al. 2009) or
mass acretion history (Chen et al. 2019). 

Observationally, the mass-tempertature scaling is found to be quite tight (with a scatter not exceeding 
20\% at most; Arnaud et al. 2005; Vikhlinin et al. 2006; Zhang et al. 2006, 2007; Lovisari et al. 2020), but past studies suffer
from several limitations in the  samples used or in the analyses applied.
All, or almost all,
samples used to  measure mass-temperature scaling studied so 
far are either a collection of objects or are X-ray- or SZ-selected. 
These samples lacks clusters of low surface brightness (Pacaud et al. 2007;  Andreon \& Moretti 2011; 
Eckert et al. 2011; Planck Collaboration 2011, 2012; Andreon et al. 2016, 2022; O'Sullivan et al. 2017, Pearson et al. 2017, Xu et al. 2018; Orlowski-Scherer et al. 2021). Andreon et al. (2022) shows that cluster X-ray faint for their mass, or equivalently
of low surface brightness are underrepresented by a factor of 10 in X-ray surveys and are entirely missing in SZ surveys.
If the missed population obeys a different scaling law
than the observed one (e.g., has a lower
intercept, a different slope, or a wider scatter), it is not possible to
recover the parameters of the entire population when no or too few examples of the missing one are present in the 
sample. 
Therefore results based on such samples, when corrected for sample
selection
(e.g. Giles et al. 2016; Liu et al. 2016; Chiu et al. 2022), necessarely implicitly assume that 
the missed
population obeys the same scaling relation as the one observed. 
Furthermore, very often hydrostatic masses are used 
(e.g. Arnaud et al. 2005, Viklinin et al. 2007), which restrict the sample 
to relaxed clusters only because the 
hydrostatic equilibrium might not be correct for unrelaxed
clusters. In simulations, un-relaxed clusters show a different scatter (Nagai 2007b),
further suggesting
that the mass-temerature relation based on relaxed clusters is biased. 

The soundness of the analysis is often hampered by the fitting methods used. 
Many works in the past have used simplified fitting methods that did  
not account for the sample selection function and the non-constant mass function (see
Andreon \& Hurn 2013 for a review on statistical issues 
related to scaling relations). Furthermore,
hydrostatic masses (Sarazin 1988), widely used for mass-temperature scaling relations, are, by definition, derived directly from  $T$ and therefore they cannot be considered independent from it, in spite of what most analyses performed thus far assume.  This introduces possible biases, the most obvious
of which is an underestimate of the scatter around the mean relation because the covariance in the errors is neglected
(see also Zhang et
al. 2006). Masses estimated from gas masses,
gas fraction, or $Y_X$ (Kravtsov et al. 2006) have similar shortcomings, as discussed in Sec.~3.
When the  clusters considered have
different redshifts, as it is often the case, 
there is a degeneracy between the evolution and slope of the scaling relation,
as discussed in detail in Sec.~4. 

A sample selected independently\footnote{More correctly, the selection should be independent at fixed mass.} of the intracluster medium being studied and with known masses 
would allow us to directly see where the population of clusters missed by X-ray and SZ-selected
samples falls in the mass-temperature diagram, and therefore to unveil weather this hard-to-detect population obeys the same mass-temperature scaling relation as the one determined from current samples. The
X-ray Unbiased Cluster Sample (XUCS, Andreon et al. 2016) is ideal for this purpose 
since its selection is independent of the intracluster medium, it contains a sizeable fraction of these low X-ray surface brightness clusters and masses are not measured from the X-ray photons used
to infer temperature, therefore making  the estimates of temperature and mass independent.

Throughout this paper, we assume $\Omega_M=0.3$, $\Omega_\Lambda=0.7$, and $H_0=70$ km s$^{-1}$ Mpc$^{-1}$. 
Results of stochastic computations are given in the form $x\pm y$, where $x$ and $y$ are 
the posterior mean and standard deviation. The latter also
corresponds to 68\% uncertainties because we only summarize
posteriors close to Gaussian in this way. All logarithms are in base 10. 

\section{The sample and the measurements}

The XUCS sample consists of 34 clusters in the very
nearby universe ($0.050 < z < 0.135$) selected from
the SDSS spectroscopic survey using more than 50
concordant redshifts whithin 1 Mpc and a velocity dispersion of
members $\sigma_v>500$ km/s (see Paper I and Paper II). They are in regions 
of low Galactic absorption, by design. 
There is no X-ray selection in the sample in the sense that the
probability of inclusion of a cluster is independent of
its X-ray luminosity, or any X-ray property, and no cluster is added/removed on the basis of its
X-ray luminosity. All clusters were followed-up in the X-ray band 
with Swift,
except for a few with adequate XMM-Newton or Chandra data in the archives.

Caustic masses (Diaferio \& Geller 1997 and later works), that have the advantage of
not relying on the dynamical equilibrium hypothesis, were derived  within $r_{200}$ in
Paper I, using
more than 100 galaxy velocities, on average, per cluster,
and then converted into masses within $r_{500}$, $M_{500}$, assuming a
Navarro et al. (1997) profile with concentration five (using a value of three would
not have changed the results, see Paper I). The average mass error is 0.14
dex. In our sample, we found that caustic masses are consistent with dynamical and weak-lensing
masses (Paper I and III), and for a single cluster with a deep X-ray follow-up, also with
the hydrostatic mass (Paper IV). In other samples, caustic masses are, in general, found to
agree with weak-lensing masses and hydrostatic masses (Geller et al. 2013, Hoekstra et al. 2015, Maughan et al. 2016). 

\begin{table}
\caption{Results of the analysis.}
\begin{tabular}{lrrrrr}
\hline\hline
Id  & \multicolumn{1}{c}{RA} & Dec & \multicolumn{1}{c}{z} & $T_X(r<300)$  \\
   & \multicolumn{2}{c}{(J2000)}& &  \multicolumn{1}{c}{keV} \\
\hline
 CL1001 & 208.2560 &  5.1340  & 0.079  & $5.7^{0.4}_{-0.3}$   \\
 CL1009 & 198.0567 & -0.9744  & 0.085  & $3.1^{0.2}_{-0.2}$  \\
 CL1011 & 227.1073 & -0.2663  & 0.091  & $2.9^{0.8}_{-0.6}$   \\
 CL1014 & 175.2992 &  5.7350  & 0.098  & $3.3^{0.4}_{-0.4}$   \\
 CL1015 & 182.5701 &  5.3860  & 0.077  & $3.9^{0.2}_{-0.2}$   \\
 CL1018 & 214.3980 &  2.0530  & 0.054  & $2.4^{0.6}_{-0.4}$   \\
 CL1020 & 176.0284 &  5.7980  & 0.103  & $5.0^{0.7}_{-0.6}$   \\
 CL1030 & 206.1648 &  2.8600  & 0.078  & $1.6^{0.2}_{-0.2}$   \\ 
 CL1033 & 167.7473 &  1.1280  & 0.097  & $2.8^{0.6}_{-0.5}$   \\
 CL1038 & 179.3788 &  5.0980  & 0.076  & $4.7^{0.9}_{-0.7}$   \\
 CL1039 & 228.8088 &  4.3860  & 0.098  & $6.0^{0.4}_{-0.3}$  \\
 CL1041 & 194.6729 & -1.7610  & 0.084  & $6.2^{0.1}_{-0.1}$   \\
 CL1047 & 229.1838 & -0.9693  & 0.118  & $7.1^{1.9}_{-1.4}$  \\
        & 	   &	      &        & $5.1^{0.1}_{-0.1}$ \\
 CL1052 & 195.7191 & -2.5160  & 0.083  & $3.5^{0.1}_{-0.1}$  \\
 CL1067 & 212.0220 &  5.4180  & 0.088  & $3.6^{0.7}_{-0.4}$   \\
 CL1073 & 170.7265 &  1.1140  & 0.074  & $2.7^{0.4}_{-0.2}$   \\
        & 	   &	      &        & $3.0^{0.6}_{-0.5}$   \\
 CL1120 & 188.6107 &  4.0560  & 0.085  & $1.2^{0.2}_{-0.2}$ \\
 CL1132 & 195.1427 & -2.1340  & 0.085  & $1.4^{0.2}_{-0.2}$   \\
 CL1209 & 149.1609 & -0.3580  & 0.087  & $2.1^{0.1}_{-0.1}$  \\
 CL2007 & 46.5723  & -0.1400  & 0.109  & $1.6^{0.6}_{-0.2}$   \\
 CL2010 & 29.0706  &  1.0510  & 0.080  & $2.6^{0.4}_{-0.3}$   \\
 CL2015 & 13.9663  & -9.9860  & 0.055  & $3.7^{1.3}_{-0.9}$   \\
 CL2045 & 22.8872  &  0.5560  & 0.079  & $2.7^{0.5}_{-0.3}$   \\
 CL3000 & 163.4024 & 54.8700  & 0.072  & $2.5^{0.2}_{-0.2}$   \\    
 CL3009 & 136.9768 & 52.7900  & 0.099  & $1.5^{0.2}_{-0.2}$   \\
 CL3013 & 173.3113 & 66.3800  & 0.115  & $5.3^{0.7}_{-0.4}$   \\
 CL3020 & 232.3110 & 52.8600  & 0.073  & $3.6^{0.7}_{-0.6}$   \\
 CL3023 & 122.5355 & 35.2800  & 0.084  & $4.5^{1.4}_{-1.1}$   \\
 CL3030 & 126.3710 & 47.1300  & 0.127  & $7.0^{0.8}_{-0.6}$   \\
 CL3046 & 164.5986 & 56.7900  & 0.135  & $10.8^{2.0}_{-1.2}$  \\
        &  &   &  & $10.7^{1.2}_{-0.8}$ \\
 CL3049 & 203.2638 & 60.1200  & 0.072  & $2.9^{0.5}_{-0.3}$   \\
 CL3053 & 160.2543 & 58.2900  & 0.073  & $2.5^{1.8}_{-0.8}$   \\
\hline
\end{tabular}
\hfill \break 
Three clusters have been observed by two telescopes and therefore are listed twice.
Coordinates and redshifts are from Paper I.
\label{tab1}
\end{table}

The X-ray data reduction is described in Paper I. Briefly, we reduced
the X-ray data (mostly Swift, but also Chandra and XMM-Newton, with three clusters observed
by Swift and either XMM-Newton or Chandra) using the standard data reduction procedures
(Moretti et al. 2009; XMMSAS 2 or CIAO 3). 
In paper I we derived 
$L_{500}$,
the cluster X-ray luminosity within $r_{500}$,  and its core-excised version
$L_{500,ce}$, calculated excluding radii smaller than $0.15r_{500}$.  
To have a glimpse on the spread of the quality of the determinations,
the median X-ray luminosity error is 0.04 dex, the error interquartile range is (0.02,0.06) dex, the worst
determination has 0.15 dex error.  We took the position of the brightest cluster galaxy closest to the
X-ray peak as the cluster center.

One cluster, CL1022, is bimodal and therefore measuring its central  temperature  is meaningless.
Another cluster, CL2081, is too faint to derive a robust measurement of the temperature.
For these reasons, these two clusters were dropped from the sample that we will use in next sections. 

\begin{figure}
\centerline{\includegraphics[trim=0 30 0 0,width=7truecm,angle=-90]{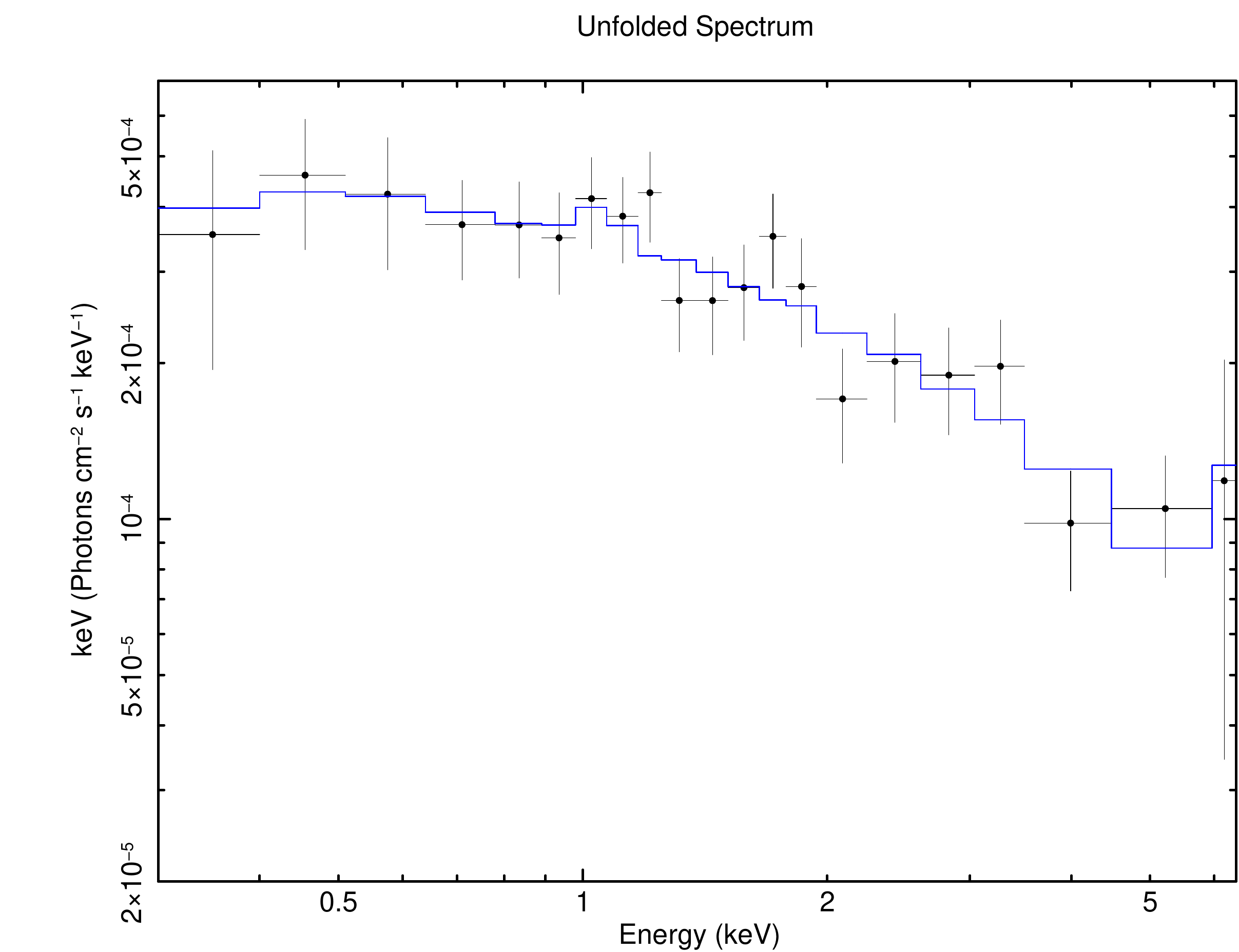}}
\caption[h]{Example spectrum. The figure shows the
spectrum of the cluster CL1038, which has median value of net photons in the spectroscopic aperture. 
}
\label{fig:spec}
\end{figure}

Here, we measure the X-ray temperature within a 300 kpc aperture following the same procedure
adopted for temperature measurements in a larger aperture in Paper II:
first, we extracted photons in [0.3-7] keV band in two regions, $r<300$ kpc 
and the annulus between $r/r_{500} = 0.7$ and  the 50\%
exposure boundary. As discussed in Paper II, the background annulus is far enough not to bias
the X-ray temperature determination.
The source spectrum was fit with an absorbed APEC (Smith
et al. 2001) plasma model, with the absorbing column fixed at
the Galactic value (Dickey \& Lockman 1990), the metal 
abundance fixed at 0.3 relative to solar, and the redshift of the plasma
model fixed at the optical redshift. The fit accounts for variations in
exposure time, excised regions, etc. The spectrum
was grouped to contain a minimum of five counts per bin and
the source and background data were fitted within the XSPEC (Arnaud 1996)
spectral package using the modified C-statistic (also called W-statistic in XSPEC). 
The median number of net photons within the aperture is 712 (interquartile range $=$(404,2365)).
Cluster photons are, on average, 60\% of all photons in the aperture because of the low background of XRT and because XRT is used for almost all clusters.
The median temperature error is 15\%.  
The error
interquartile range is (9,20)\%.
We find consistent temperature values for the three clusters observed with XRT and 
Chandra or XMM-Newton (with a difference of $<1.4\sigma$).

Table~\ref{tab1} lists derived temperatures. An example of the ``average" quality of the spectral data is given in
Fig.~\ref{fig:spec}, where we show the spectrum of CL1038 that contains the median value of net photons. 

\section{Results}

\subsection{$L_X-M$}

\begin{figure}
\centerline{\includegraphics[width=9truecm]{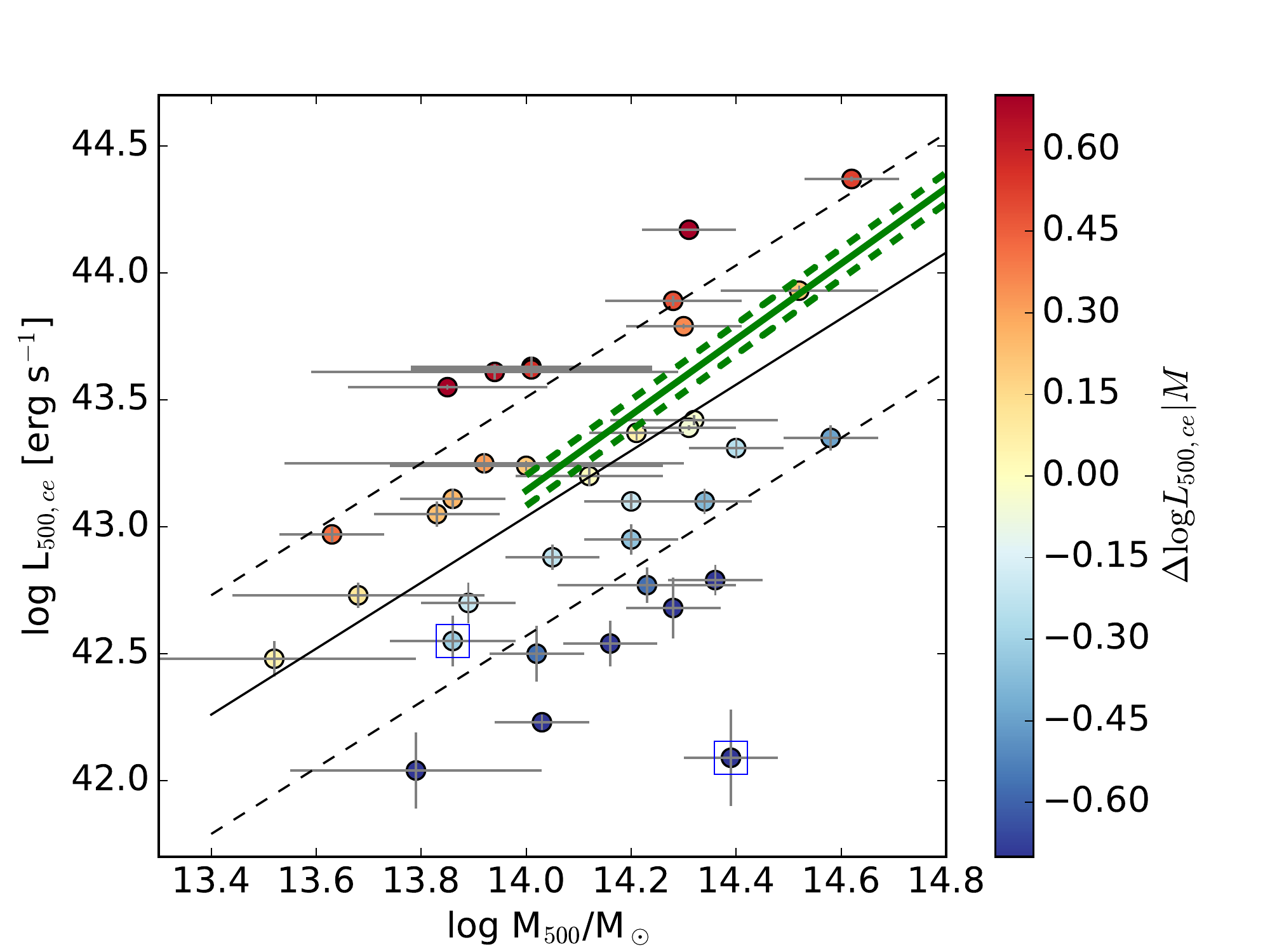}}
\caption[h]{The sample. Core-excised luminosity within $r_{500}$ in the [0.5-2] band vs $M_{500}$
color-coded by offset from the mean 
relation. The two objects excluded by the current analysis 
are marked by squares. The blue solid line and the dashed  corridor indicate the mean relation and
the $\pm 1\sigma_\text{intr}$ regions derived in Paper I and II. Three clusters (observed by
two telescopes) appear twice in the figure, although distinguishing these duplicates in the plot can be sometime challenging. 
The green solid line and the dashed corridor indicate the mean relation and
the $\pm 1\sigma_\text{intr}$ regions derived for the X-ray selected REXCESS sample.
}
\label{fig:LxMoffcod}
\end{figure}

The 
XUCS $L_X-M$ relation was presented and
discussed in Paper I. In Fig.~\ref{fig:LxMoffcod} we show it 
again, here color coded by the offset from the
mean relation. The Figure also shows 
the mean relation and
the $\pm 1\sigma_\text{intr}$ regions for the X-ray selected REXCESS sample (Bohringer et al. 2007)
with the green solid line and the dashed corridor.  The 
XUCS sample (data points) occupies a much wider portion of the $L_X-M$ plane than REXCESS, as discussed in Paper I.
Blue points represent 
under-luminous clusters (faint for their mass), red ones are over-luminous.
The offset can also be
read as a difference in the average surface brightness within $r_{500}$: clusters 1 dex brighter in $L_{500,ce}$
are also 1 dex brighter in average surface brightness compared to clusters with the same $r_{500}$ (or $M_{500}$),
by definition. A pair of such clusters is shown in Fig.~2 and 4 of Paper I. 
Paper II shows these differences to be related to
gas fraction: clusters that are X-ray faint for their mass are also gas poor. 
Furthermore,
once X-ray luminosity is corrected by gas fraction, the X-ray luminosity vs mass scaling become
almost scatterless (Paper II). In other terms, the color coding  indicates equivalently a difference
in X-ray luminosity, in mean surface brightness within $r_{500}$, or in gas fraction at fixed mass.
In the Figures we use $\Delta \log L_{500,ce} |M$ (the vertical
bar reads ``at") which is more directly related to the data, but we could have equally used $\Delta \log f_{gas} |M$ or
$\Delta \log \mu_{r_{500}} |M$. 
The two clusters dropped from further analysis 
are identified with a square.

\subsection{$T-M$}

\begin{figure}
\centerline{\includegraphics[width=9truecm]{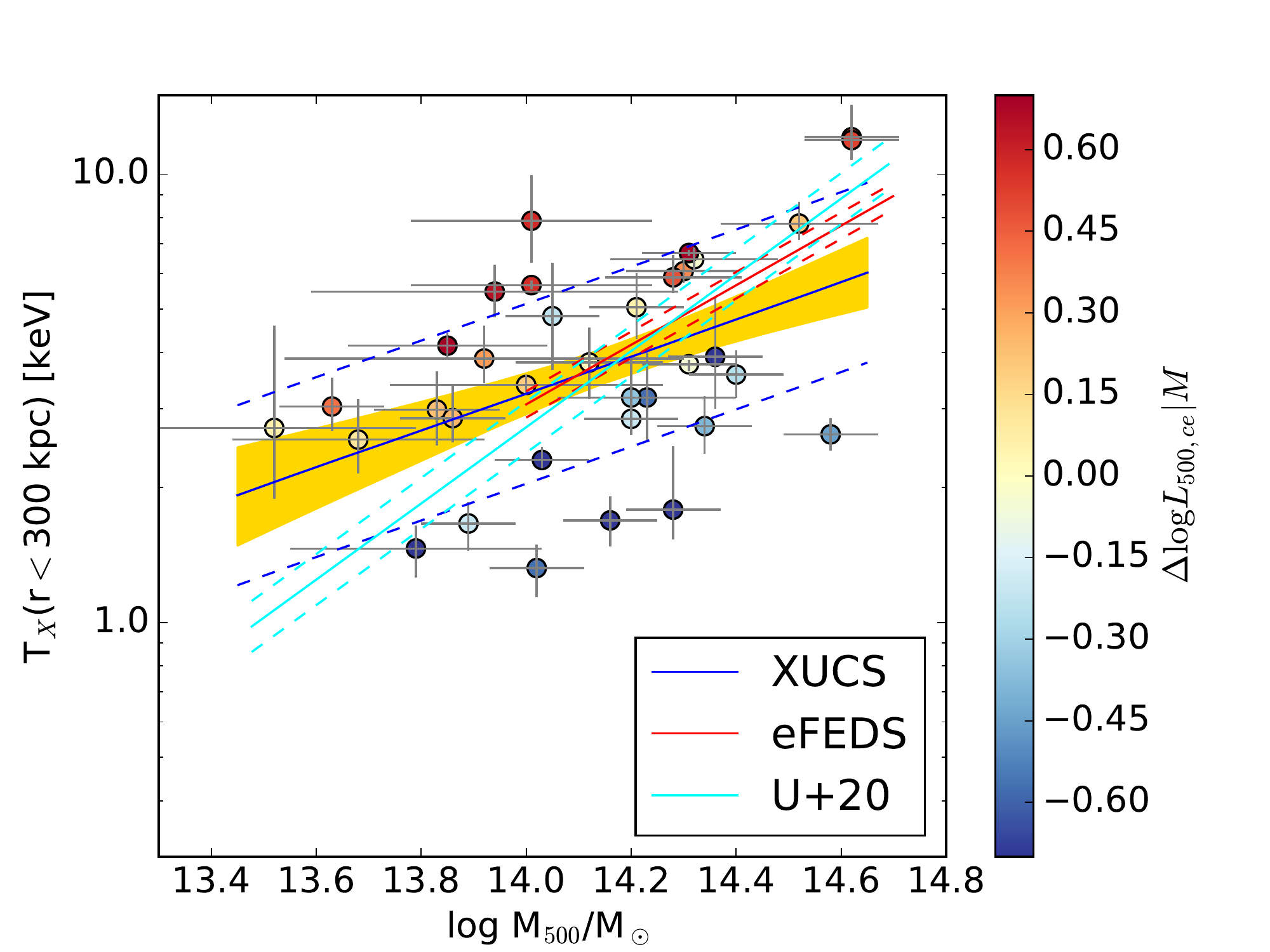}}
\caption[h]{Temperature-Mass relation
color-coded by the offset from the mean $L_{X,ce}-M$ relation.
The mean model and its 
68\% error are shown as a blue solid line and yellow shading, respectively.
Clusters faint for their $M_{500}$ (blue points) have a systematically cooler temperature for their mass. 
Three clusters (observed by two telescopes) appear twice in the figure.  Fits from the literature are also shown, with 68\% corridors indicated by dashed lines, which severely  underestimate
the spread observed in the XUCS sample.
}
\label{fig:TMoffcod}
\end{figure}

Fig.~\ref{fig:TMoffcod} shows the XUCS sample in the mass-temperature plane. In place of the common tight relation, the sample shows a relatively large scatter
and shows that the location of a cluster in the plot is closely related to  
the offset from the $L_{500,ce}-M$ mean relation: blue points
are systematically below red points. As will be illustrated below, this occurs because
$T$ is directly related to $L_X$ and only indirectly to $M$.

To derive the scatter, we adopt
Gaussian errors in $\log M_{500}$ and lognormal errors on $T_X$ (introduced in Andreon 2012 and now
widely adopted) and a Gaussian intrinsic scatter in $\log T|M$. For the three clusters with observations
with two telescopes, we use the measure with smaller errors.
We adopt weak priors for the parameters 
and a uniform prior in $\log M$. Since the sample is not X-ray selected, we don't
need to model the X-ray selection function and to make hypothesis
on the amplitude and spread of the unseen population.
The numerical implementation of the code is given
in Sec.~8.4 of Andreon \& Weaver (2015). We found
\begin{equation}
\log T = (0.42 \pm 0.14) (\log M_{500} -14.2) + 0.59 \pm0.04
\end{equation}
where $T$ is in keV and $M_{500}$ is in $M_\odot$.
The intrinsic scatter is $0.20\pm0.03$ dex at fixed mass and
there is almost no covariance between parameters because of
our choice of pivoting masses near the middle of the data distribution (i.e. subtracting 14.2 to $\log M_{500}$).
The intrinsic scatter is maximally large since the data
scatter in $T$ is 0.22 dex.
An identical result is obtained by taking a Schechter (1976) as prior mass function\footnote{The code for the numerical implementation of the Schechter (1976) prior is given in Andreon \& Berge (2012).}, basically
because the errors on $M$ are small compared to the prior width for almost all data points. This indicates
that our results are independent of the shape of the $M$ distribution adopted for the sample.

Fig.~\ref{fig:TMoffcod} also shows two
determinations of the mass-temperature relation from the literature which use, as we do, 
non-core-excised temperatures. Both use weak-lensing masses from the Hyper Suprime-Cam 
Subaru Strategic Program survey (Aihara et al. 2018) and X-ray selected samples:
Umetsu et al (2020, cyan in the figure) 
use the XXL survey
(Pierre et al. 2016), whereas Chiu et al. (2022, red in the figure) uses the eROSITA Final Equatorial-Dept
survey (eFEDS, Liu et al. 2022). 
Umetsu et al (2020) use $T(<300)$ kpc, as we do, whereas Chiu et al. (2022) use $T_(<r_{500}$)
found by Giles et al. (2016) to be indistinguishable from $T(<300)$ kpc given the errors.
Umetsu et al (2020) and Chiu et al. (2022) find an intrinsic scatter of $0.06\pm0.05$  and $0.03^{+0.03}_{-0.01}$ dex at fixed mass, 
respectively vs $0.20\pm0.03$ dex for XUCS. In Fig.~\ref{fig:TMoffcod}
dashed corridors mark the $\pm 1 \sigma_{intr}$ region around the mean relation, which in absence of errors include
68\% of the data points. These two relations and reported scatter 
severely underestimate the scatter
observed in XUCS. This occurs because both XXL and eFEDS lack clusters of low surface brightness (the blue and cyan points)
and therefore from these data it is not possible to recover the full size of the mass-temperature distribution and to properly estimate the region  occupied by
low surface brightness clusters from the detected clusters with high surface brightness. 
The slope of the mass-temperature relation is very poorly determined in Umetsu et al. (2020) because the errors on the mass are larger than the explored mass range, while Liu et al. (2022) slope determination assumes that clusters of all brightnesses obey a brightness-independent mass-temperature scaling, which XUCS data manifestly show not to be true.
These
discrepancies emphasize the importance of using cluster samples inclusive of low surface brightness
objects and of developing fitting schemes that allows us to properly infer the spread of the population when
a part of it is severely undersampled, as in X-ray surveys.

More in general, the unprecedent large spread  observed in the XUCS sample does not show up in X-ray or SZ selected samples for two reasons.
First, as mentioned, because they rarely include low luminosity/low surface brightness clusters (namely the blue points, see Andreon et al. 2016,2022).
Second, the observed scatter is strongly reduced by assuming a scatterless scaling relation to
infer the radius $r_{500}$ in which measurements have to be performed. 
For example, in analyses where $r_{500}$ is estimated from $T$ (e.g., Giles et al. 2016), a cluster of low/high temperature for its mass will
have an estimated $r_{500}$ appropriate to put it precisely on the mean $M_{500}-T$ relation because this same relation, in the mathematically equivalent $r_{500}-T$ form, is used to infer $r_{500}$ from $T$.
With this assumption, cold clusters are forced to have
a small $r_{500}$ and therefore to be of low mass, even when 
the low temperature is related to low luminosity and surface brightness.
Similarly, if mass is estimated assuming that all clusters share
the same gas fraction (e.g. Mantz et al. 2010), it will again be underestimated for clusters of low surface brightness (gas fraction). The assumption of an average gas fraction moves clusters closer to a mean $M-T$ relation. Finally, the scatter is strongly reduced using combinations of
temperature and gas mass or gas fraction, such as $Y_X$ (Kravtsov et al. 2006), 
to estimate $r_{500}$ since, as before,
cluster with low gas fraction or cool because of their low surface brightness, will have
underestimated mass and $r_{500}$.

\begin{figure}
\centerline{\includegraphics[width=9truecm]{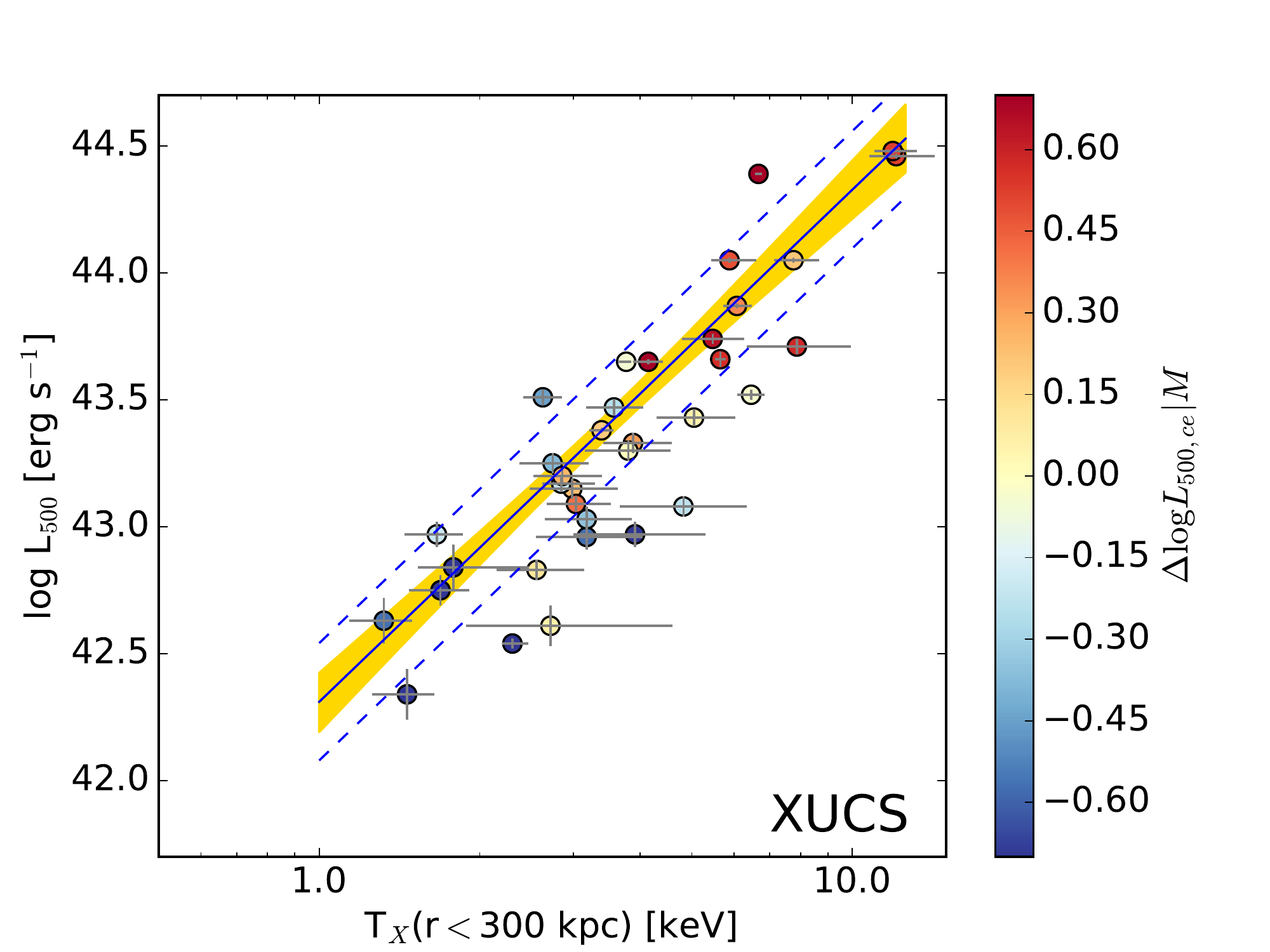}}
\caption[h]{$L_{500}-T$ relation, 
color-coded by the offset from the mean $L_{X,ce}-M$ relation. The mean model and its 
68\% error are shown as a solid line and yellow shading, respectively. 
The mean model $\pm 1\sigma_\text{intr}$ is
shown as dashed corridor. 
There is a tight scaling between $L_{500}$ and $T$, independent on cluster over/under luminosity
(blue/red points are not systematically above or below the average), see also Fig.~\ref{fig:residLxToffcod}. 
Three clusters (observed by
two telescopes) appear twice in the figure, but only the ones with smaller errors are 
used in the fit.
}
\label{fig:LxToffcod}
\end{figure}

\subsection{$L_X-T$}

Fig.~\ref{fig:LxToffcod}  shows the $L_{500}-T$ relation of XUCS clusters. To derive the  fit parameters, we adopt 
Gaussian errors in $\log L_{500}$ and lognormal errors on $T_X$ 
and a Gaussian intrinsic scatter in $\log L_{500}|M$. 
Assuming weak priors and an uniform prior on $\log T$, we found
\begin{equation}
\log L_{500} = (2.02 \pm 0.22) \log (T/3) + 43.27 \pm0.05
\end{equation}
where $T$ is in keV and $L_{500}$ is in \ergs,
with an intrinsic scatter of $0.23\pm0.04$ dex and almost no covariance between parameters due to  
our choice of pivoting $T$ near the middle of the data distribution, 3 keV. The intrinsic scatter is
much smaller than the range explored by the data (two dex, in the ordinate), unlike the mass-temperature relation
were the intrinsic scatter is as large as allowed by the data scatter.
An identical result is obtained by taking a Schechter (1976) prior mass converted in $T$ using the
scaling relation in equation 1 inclusive of scatter, basically
because the errors on $T$ are small compared to the prior width for almost all data points. This indicates
that our results are independent of the shape of the $T$ distribution adopted for the sample.

\begin{figure}
\centerline{\includegraphics[width=9truecm]{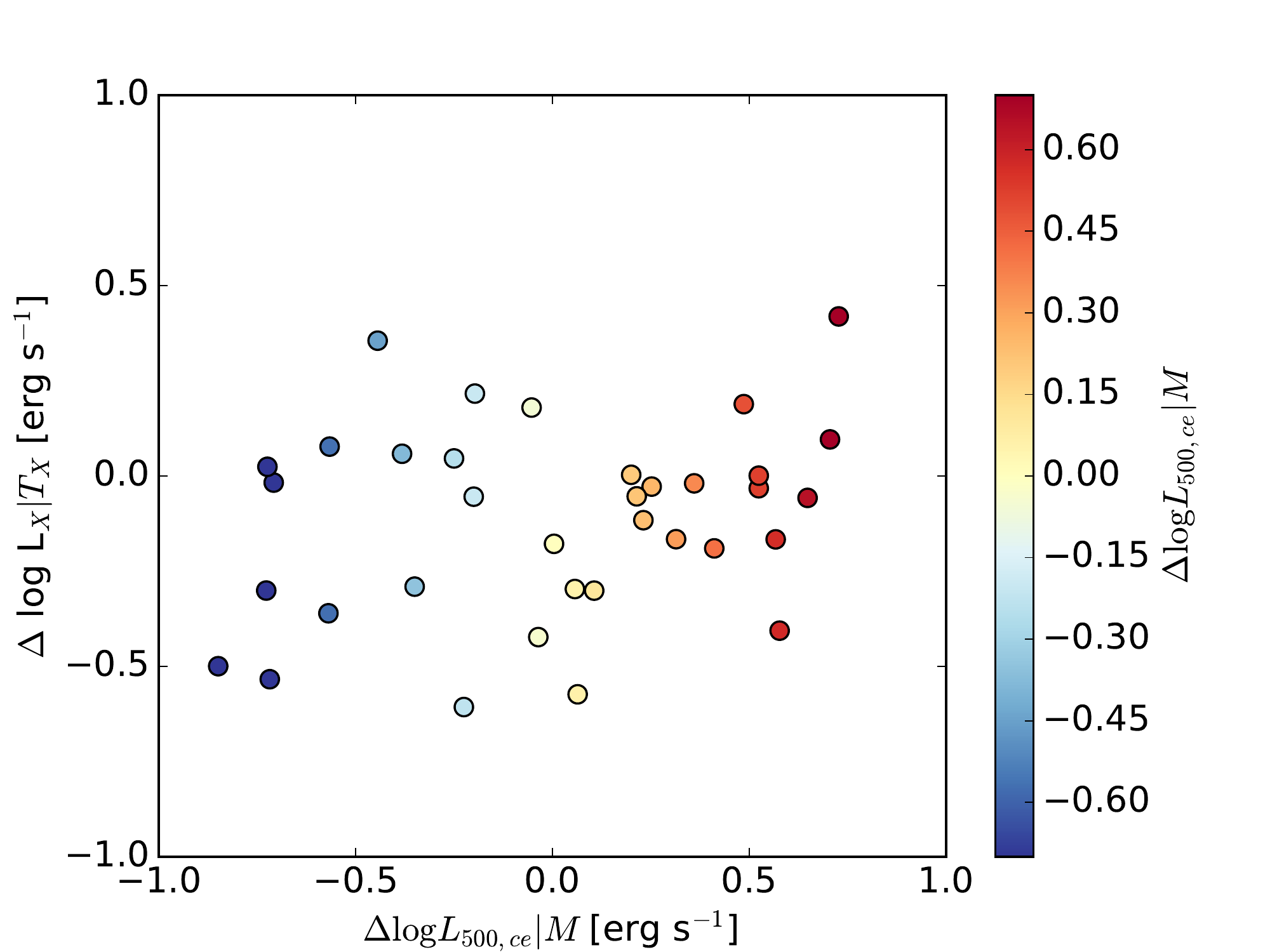}}
\caption[h]{Residuals from the $L_{500}-T$ relation vs residuals from the $L_{500,ce}-M$
relation color-coded by offset from the mean $L_{X,ce}-M$ relation. Clusters faint/bright
at a given mass are not systematically hot/cold.  Three clusters (observed by
two telescopes) appear twice in the figure.
}
\label{fig:residLxToffcod}
\end{figure}

Blue/red points are not systematically above, or below, the best fit. Therefore
clusters have a temperature appropriate for their X-ray luminosity, regardless of their position on the $L_{X,ce}-M$ plane, i.e. of whether they are under/over luminous, of low/high surface brightness
or with a small/large gas fraction.  This
can be better seen in Fig.~\ref{fig:residLxToffcod} which shows the residuals from the $L_{500}-T$ 
and the $L_{500,ce}-M$ relations.

We interpret the tighter trend between $T$ and $L_{500}$ than $T$ and $M_{500}$ as due to 
the fact that
X-ray luminosity and temperature are quantities dominated by photons
coming from small radii (in particular for $T$ we used 300 kpc) and
strongly affected by gas physics (e.g. shocks, cool-cores),
while mass is dominated by dark matter at large radii:
for example in a NFW profile with 
$c_{500}=5$ and $r_{500}=1$ Mpc, 70 \% of the mass is at $r>300$ kpc. Thermodynamic properties
at small radii are not much affected by the mass distribution at large radii, as even simple
numerical experiments show,
making temperature little affected by the amount of mass at large radii.  Furthermore,
during major, intermediate, and minor mergers, evolution moves simulated clusters in the $L_X-T$ plane mostly along the
$L_X-T$ relation (e.g., Richter \& Sarazin 2001; Poole et al. 2007).
Therefore these events do not add scatter in the $L_{500}-T$ plane, while they might in the $L_{500}-M$ plane,
for example,
differences in mass accretion histories may contribute
to the observed large scatter in the temperature-mass plane.  
To summarize, while in a simplified universe clusters
are close to isothermal spheres in hydrostatic equilibrium with a scatter-less scaling between mass and temperature,  
the observed mass-temperature plot reminds us that 
what happens to the gas at small radii is not felt by dark matter at large
radii and neither directly affected by it.

The analysis of deep follow-up data of one single X-ray faint cluster (Paper IV)
already highlighted this detachment between local and global quantities, at least in that single object,
and showed that this extends to other local/global quantities, such as central entropy and 
integrated pressure.

Fig.~\ref{fig:LxToffcod} suggests that clusters that are X-ray faint for their mass, or of low surface brightness 
within $r_{500}$, cannot be recognized in a $L_{500}-T$ diagram: they are not systematically offset from the mean relation, and therefore cannot be singled out in  this kind of
analysis. This indicates that  the $L_{500}-T$ relation derived  for  X-ray selected
samples is robust because the missed population (of low surface brightness) obeys the same $L_{500}-T$ of the detected
population, as these analyses assume. Therefore, the lack of objects of low surface brightness in these samples  does
not introduce a (strong) bias in $L_{500}-T$ analyses. It will, instead, in $M-T$ analyses as
obvious from Fig.~\ref{fig:TMoffcod}.

\section{Discussion}

The $L_{500}-M$ relation derived for X-ray selected samples is biased high because cluster faint for their mass are underrepresented (Paper I). One may wonder if a similar bias might exist for the $L_{500}-T$ scaling as well. Our X-ray unbiased sample already indicates this is not the case because this scaling relation does not depend
on the offset from the $L_{500,ce}-M$ scaling law (sec.~3). Nevertheless, a direct comparison 
with X-ray selected samples would be reassuring. 

\begin{figure}
\centerline{\includegraphics[width=9truecm]{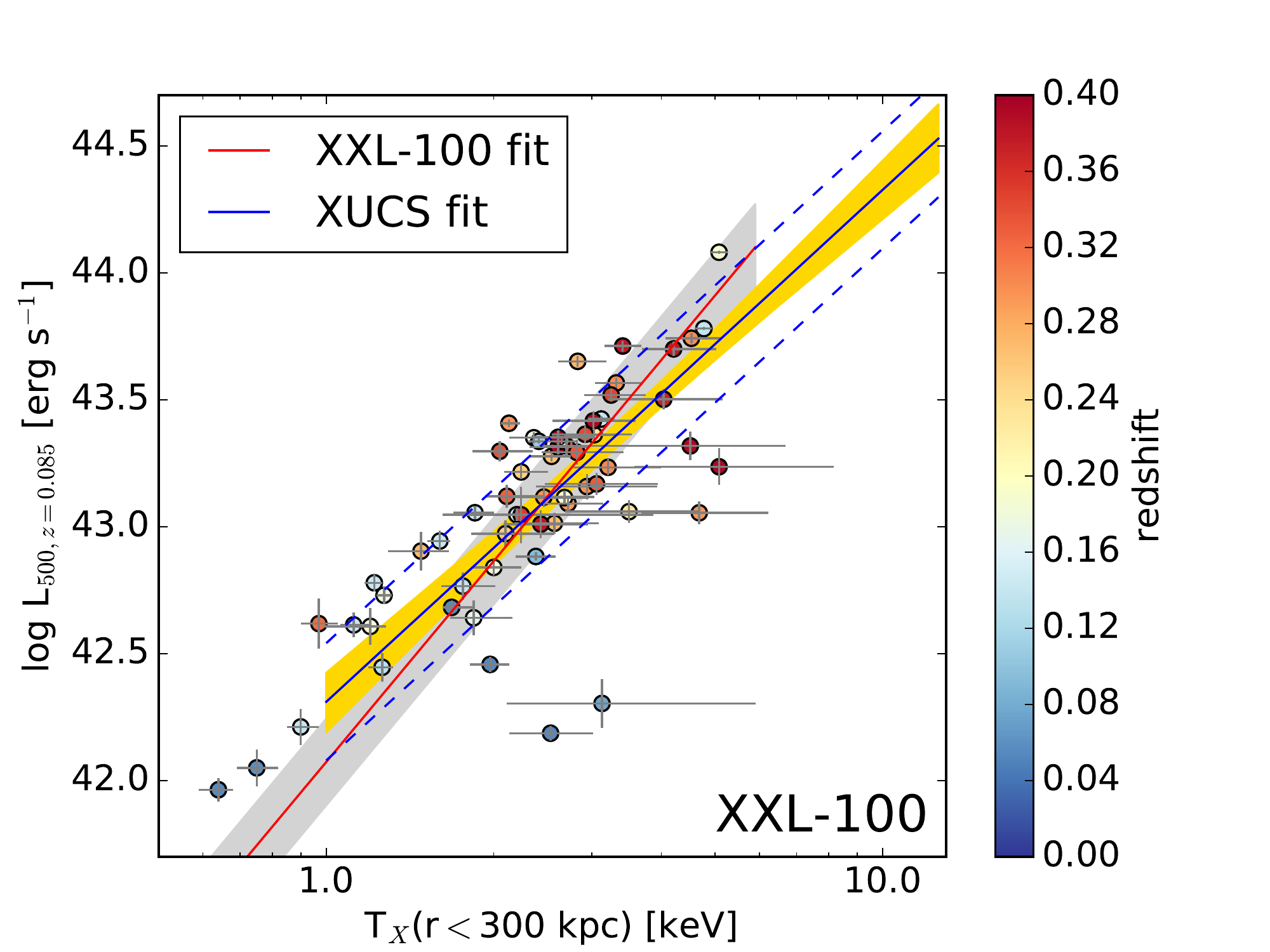}}
\caption[h]{XXL-100 $L_{500}-T$ relation
color-coded by redshift,  corrected for evolution  at the median redshift of XUCS. The two points
with $\log L_{500}\sim42.2$ and $T\sim3$ kT,
likely have problematic measurements (see Giles et al. 2016 and text).
The red solied line is the best fit to the whole XXL-100 sample,
while the gray shading marks the mean $\pm 1 \sigma_\text{intr}$ corridor, as derived by Giles et al. (2016).
The blue solid line, the yellow shading and the dashed lines mark the fit,
uncertainty and $\pm 1 \sigma_\text{intr}$ corridor of the XUCS sample. The two fits are
consistent, although ours is derived with fewer assumptions and without the approximations
assumed in XXL-100. Only $z<0.4$ XXL-100 clusters are shown,
to focus on those with lower evolutionary corrections.
}
\label{fig:XXL100LxT}
\end{figure}

Fig.~\ref{fig:XXL100LxT} shows the $L_{500}-T$ plot of the XXL-100 sample, color-coded by
redshift, and corrected for evolution to the median redshift of the XUCS sample,
$z=0.085$, according
to the best fit evolution computed by Giles et al. (2016). 
The two  points with $kT\sim3$ keV and $\log L_{500}\sim42.2$ \ergs that stand out of the mean relation  are likely to have measurement problems
(Giles et al. 2016) and one of them had $kT=0.6^{+0.2}_{-0.1}$ in an earlier analysis (Pierre et al. 2006). 

XXL-100 data points are in broad agreement with XUCS fit (Fig.~\ref{fig:XXL100LxT}).
The Giles et al. (2016) XXL-100 best-fit relation (slope, intercept, and scatter) agrees with the one derived for XUCS within the errors. However, a later, very similar, analysis adopting similar assumptions for an enlarged sample by the same
team (Adami et al. 2018) prefers a steeper slope inconsistent with XUCS one ($3.17\pm0.16$ vs XUCS $2.02\pm0.22$, about $4\sigma$ away, and XXL-100 $2.63\pm0.15$)\footnote{Since all three works use pivot masses, there is almost no
covariance between intercept and slope. Therefore, scaling relations can be compared via the slope alone. 
slope-intercept plane.
}. A closer
look at the data and assumptions done to derive the best fit parameters is therefore in order. 

As most X-ray selected samples, larger values of 
X-ray luminosity are found at large redshifts because very bright
clusters are rare and therefore are missing in the low volume of the local Universe. Since $T$ also increases with $L_{500}$, then the $L_{500}-T$ relation of X-ray
selected samples displays a redshift gradient (see color coding in Fig.~\ref{fig:XXL100LxT}): 
clusters in the top-right of the $L_{500}-T$ diagram
are the most distant ones, while those in the bottom-left corner are in the 
very nearby Universe. The effect would be more evident than seen in Fig.~\ref{fig:XXL100LxT}
also considering XXL-100 $z>0.4$ clusters because they all fall in the top-right part of the plot.
Therefore, the $L_{500}-T$ slope and evolution are largely collinear (degenerate),
i.e. the $L_{500}-T$ in such samples is equally well fitted by allowing larger slopes
and smaller evolutions (as already reported for the $Y_{SZ}$ vs mass relation
in Andreon 2014). Depending on the evolution parameter, the relation corrected for evolutionary effects
rotates, and it
does so hinged at the low-$L_{500}$ low-$T$ corner, since these clusters are in the nearby Universe and therefore
are not affected by evolutionary corrections. Since 
XUCS clusters are instead all in a narrow redshift range, $\Delta z =0.085$, their
differential evolution across the redshift range is negligible and the slope
is not collinear with evolution. 
The collinearity explains the wide range of inferred slopes.

\begin{figure}
\centerline{\includegraphics[width=8truecm]{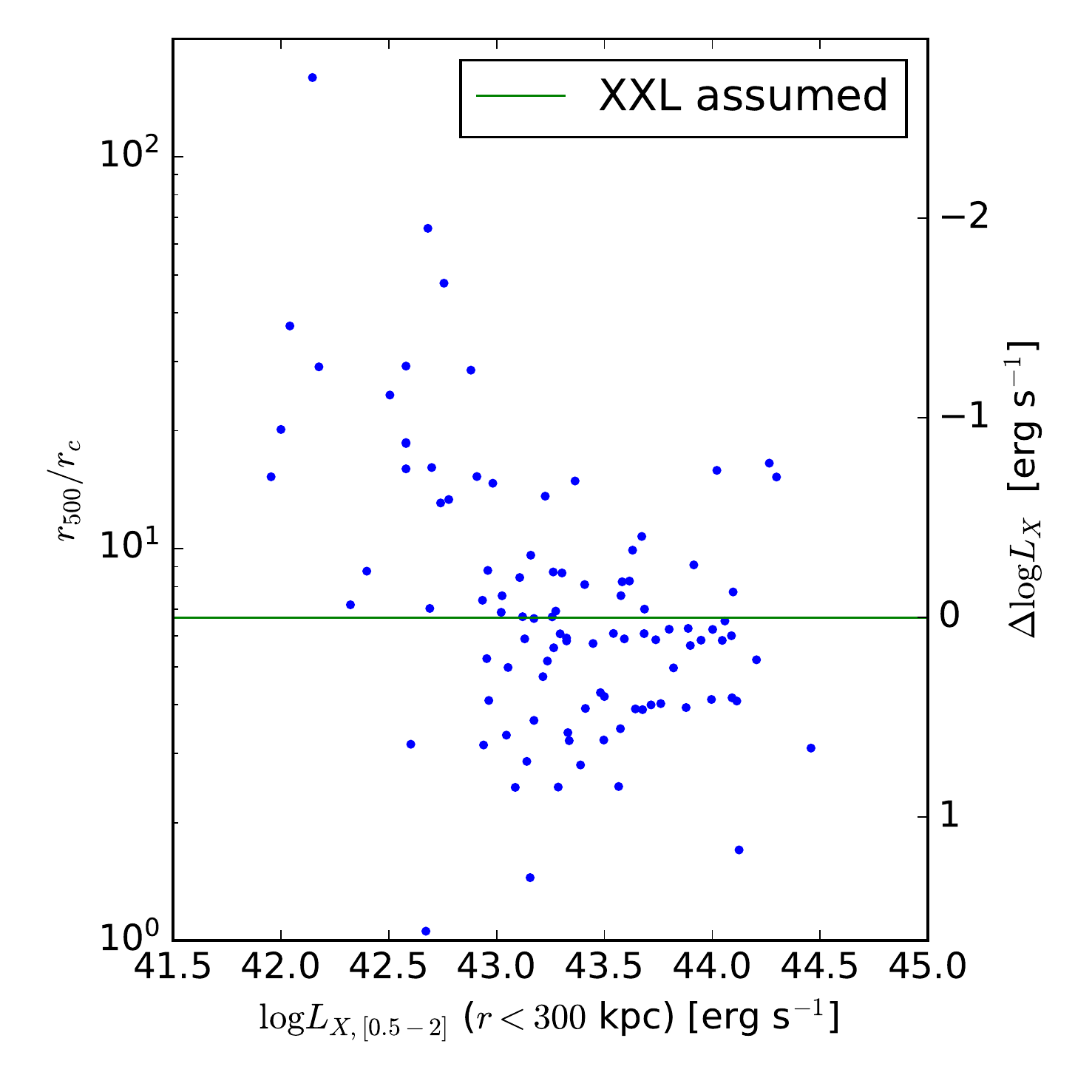}}
\caption[h]{Assumed (green line) and observed (points) $r_{500}/r_c$ 
used by XXL-100 to compute the luminosity within $r_{500}$ from the luminosity
within 300 kpc plotted in abscissa, and cluster detection probability. 
The right ordinates shows the luminosity offset
for a cluster with $r_{500}=1$ Mpc on a logarithm scale. On average, the offset is
$\pm 0.5$ dex, i.e. a factor of three. See text.
}
\label{fig:XXLextraopol}
\end{figure}

However, the collinearity does not explain why derived slopes are inconsistent,
forcing us to look for sources of error underestimation that, if accounted for, would make measurements consistent.
In their analysis, Giles et al. (2016) and Adam et al. (2018) make two assumptions, 
which are not included in their error budget. The first one is related to 
the missed population: basically, both works assume
that missed clusters
(those of low surface brightness) obey to the same $L_{500}-T$ scaling of the observed population, which
our work shows for the first time to be close to be true. Second,  as also noted by the authors, they assume that all clusters  have $r_c=0.15 r_{500}$ with no 
scatter. This assumption is made twice, both to extrapolate the X-ray luminosity within $r_{500}$ from the
luminosity within the 300 kpc aperture assuming a beta-model, and to compute the detection probability.
However, clusters are characterized by
a wide scatter in $r_{500}/r_c$, as shown by  Fig.~\ref{fig:XXLextraopol} for
the clusters in XXL-100. The left ordinate shows the ratio of published $r_{500}$ and $r_c$ values.
The scatter between assumed and measured core radius is 0.27 dex (computed from the median absolute
deviation to be tolerant to outliers), i.e. about a factor of 2. Figure 9 of Pacaud et al. (2016) shows that for clusters close to threshold of the bright
sample studied by XXL-100, detectability is a strong function of the core radius and the
sky coverage changes by more than a factor of 10 when $r_c$ is doubled. 
The right ordinate shows the ratio between the luminosity calculated assuming a fixed ratio $r_{500}/r_c$ as in XXL, and the actual  
luminosity using the $r_c$ from the fit with a  $\beta$ model when $r_{500}=1$ Mpc. 
The scatter between the luminosity calculated with the two methods is 
$\Delta \log L_{500}=0.5$ dex (i.e. a factor of 3 brighter/fainter than correct). 

Since Giles et al. (2016) and Adami et al. (2018) $L_{500}-T$ do not take these approximations 
into account in the error budget, the errors quoted are   
underestimated, 
resulting in an inconsistency between the three slope estimates.
Since XUCS sample: a) is free from the slope-evolution degeneracy; b) does not miss clusters and therefore
does not make hypotheses on the $L_{500}-T$ of the unseen population; c) does not assume
a core radius to determine the cluster flux; - 
it can be used as reference and as a profitable
prior for future analysis of X-ray samples 
to break the slope-evolutionary
degeneracy.

Ge et al. (2019) also found 
similarity between the $L_X-T$ relations of optically- and X-ray selected samples. Their data however lack 
a reliable mass proxy not allowing them to, for example, address the $M-T$ relation, or to look for
dependencies from the offset from the $L_X-M$ relation that our data allow. 
Based on the analysis of the XUCS sample, we suggest a different interpretation of the observed $L_X-T$ relation: while Ge et al. (2019) believe that low temperatures observed for rich clusters are due
to failures of their adopted mass proxy, and that therefore 
they would disappear with proper mass proxy,
our sample shows that mass is less directly related to temperature than temperature is to X-ray luminosity,
and therefore
that low temperatures are possible for clusters of large mass (and faint surface brightness) 
and this does not depend on a faulty measure of mass/richness.

\section{Conclusions}

In this work we analysed the mass-temperature and luminosity-temperature
scaling relations of a sample that includes the recently discovered class of clusters that are X-ray faint
for their mass, also known to have an average low
surface brightness 
and a low gas fraction. 
We used a sample of 32 clusters in the nearby Universe selected independently of the
intracluster medium properties, from which the first few examples were discovered in Paper I. 
Our sample can rely on mass  measurement derived without dynamical or hydrostatic
equilibrium hypothesis, with average error of 0.14 dex. Furthermore, our mass and
temperature estimates are based on different and independent sets of data.

The sample was first presented in Andreon et al (2016). Here we use the same data to derive the temperature $T$ within a 300 kpc
radius aperture, which we derive with a
median error of 15\%.

The main results of this work are that
the $M_{500}-T$ relation has a large scatter, 
and the location
of a cluster in this plane depends on the cluster surface brightness.
The wider scatter around the $M-T$ relation was missed thus far 
due to two concurrent factors: it is reduced 
if the sample does not contain
low surface brightness clusters, 
therefore covering a smaller area of the $M-T$ plane, 
as in X-ray or SZ selected samples commonly studied in literature.  
Second, the determination of  
radius $r_{500}$ and mass $M_{500}$  from the temperature, gas fraction, gas mass, or $Y_X$, 
as it is commonly done in X-ray studies, also artificially reduces the scatter.
The independent estimates of mass and temperature and the presence of clusters of low
surface brightness in the XUCS sample revealed  
the full range of the cluster population
in the $M-T$ plane.

Clusters obey a tight $L_{500}-T$ scaling, independently of their brightness
within the $r_{500}$ scale, making this diagram not useful to recognize clusters of low
surface brightness. For this reason surveys that miss clusters of low surface
brightness may infer unbiased estimates of the $L_{500}-T$ scaling parameters.

We interpret the tighter relation between  $T$ and  $L_{500}$ than 
between $T$ and $M_{500}$ as due to the fact that
X-ray luminosity and temperature are more directly related to each other than with mass: these quantities are
dominated by photons
coming from small radii (in particular for $T$ we used 300 kpc) and
strongly affected by gas thermodynamics (e.g. shocks, cool-cores),
while mass is dominated by dark matter at large radii.

Since the XUCS sample is spread on a very narrow redshift interval over which
evolutionary effects are negligible and its selection function 
does not depend on X-ray properties, our determination of the
$L_X-T$ relation is free from the collinearity (degeneracy) between evolution and slope and free from hypothesis on the undetected population, both affecting current X-ray selected
samples. We derive a slope of $L_{500}-T$ relation of $2.0\pm0.2$.
Because of the reduced number of assumptions, our determination can be used both as reference and to break the above degeneracy in X-ray selected samples.

\section*{Acknowledgements}
The authors thanks the referee, Florence Durrett, Charles Romero and Marguerite Pierre  for their comments on an early version of this paper that lead to an improved presentation.
S.A. acknowledges financial contribution from the agreement ASI-INAF n.2017-14-H.0 
This work has been partially supported by the ASI-INAF program I/004/11/5.

\section*{Data Availability}
All the values needed for reproduce the plots in this Paper concerning the XUCS sample 
are in Table 1 or were published in Paper I. Raw data are in the X-ray and SDSS archives.
Data for the XXL-100 samples are published as detailed in the quoted references.

{}

\bsp	
\label{lastpage}
\end{document}